\documentclass[fleqn,10pt,twocolumn]{wlscirep}
\usepackage[utf8]{inputenc}
\usepackage[T1]{fontenc}
\usepackage{bm}
\usepackage{graphicx}% Include figure files
\usepackage{dcolumn}% Align table columns on decimal point
\usepackage{bm}% bold math
\usepackage[]{hyperref}
\hypersetup{
    colorlinks,
    linkcolor={blue},%!50!black
    citecolor={red},%!50!black
    urlcolor={blue!80!black}
}
\usepackage[framemethod=TikZ]{mdframed}
\usepackage[most]{tcolorbox}
\usepackage{longtable}
\usepackage{mathcomp}
\usepackage{pifont}
\usepackage{xcolor}
\usepackage{amsthm}
\usepackage{PGF}
\usepackage{pgfpages}
\usepackage{amsmath}
\usepackage{amsfonts}
\usepackage{amssymb}
\usepackage{graphicx}
\usepackage{bm}
\usepackage{hyperref}
\usepackage{lipsum}
\usepackage{array}
\usepackage{booktabs}
\usepackage{tikz}
\usepackage{titlesec}
\usepackage{hyperref}
\hypersetup{bookmarksnumbered}
\def\be{\begin{equation}}
\def\ee{\end{equation}}

\usepackage{amsmath}
\usepackage{empheq}
\usepackage[most]{tcolorbox}
\newtcbox{\mymath}[1][]{%
    nobeforeafter, math upper, tcbox raise base,
    enhanced, colframe=blue!30!black,
    colback=blue!30, boxrule=1pt,
    #1}
\usepackage{mdframed}
\usepackage{lipsum}
\usepackage{PGF}
\usepackage{pgfkeys}
\usepackage{pgffor}
\usepackage{pgfcalendar}
\usepackage{pgfpages}
\usepackage{tikz}
\usetikzlibrary{calc}
\usepackage{array}
\usepackage{booktabs}
\usepackage{multirow}
\usepackage{latexsym}
\usepackage[english]{babel}
\usepackage[most]{tcolorbox}

\newtcolorbox{myquote}[1][]{%
    colback=black!5,
    colframe=black!5,
    notitle,
    sharp corners,
    borderline west={2pt}{0pt}{red!80!black},
    enhanced,
    breakable,
    }
\usepackage{enumitem}

{\begin{itemize}\item[]}
{\end{itemize}}
\usepackage{graphicx}
\usepackage{mathtools}
\usepackage{etex}
\newcommand\myeq{\mathrel{\stackrel{\makebox[0pt]{\mbox{\normalfont\tiny dis.}}}{=}}}
\title{
Formulating and indirectly proving the quantum fluctuations of space-time and a justification for applying fractional quantum gravity
}

\author[*]{Behzad Tajahmad}
\affil[*]{e-mail: behzadtajahmad@yahoo.com}

\begin{abstract}
It is well known that in quantum gravity, the very geometry of space and time is subject to continual fluctuation. The mathematical formulation for this old theory is still lacking. This article formulates this more than forty-year-old theory of quantum gravity, the result of which is that the equations of quantum gravity become fractional. The existence of such fluctuations is proved indirectly.
On the other hand, recent attention has been paid to fractional gravity. Although this type of gravity leads to brilliant results, we have no root and logical reason for applying it other than that it works.  
We actually simultaneously demonstrate that fractional gravity is generated by stochastic quantum fluctuations of space-time.  
For clarification, Einstein-Hilbert theory along with a scalar field is investigated in deformed and non-deformed minisuperspaces. The results illustrate a transition from decelerated to accelerated expansion (late-time-accelerated expansion).
\end{abstract}
\begin{document}

\flushbottom
\maketitle

\thispagestyle{empty}

%\noindent \textbf{Key points:} Please suggest $\sim 5$ key points, which should be single-sentence bullet points that summarize the article and remind readers of the take-home messages. An example of key points can be found at \url{https://www.nature.com/articles/s42254-018-0001-7#Abs3}

%\noindent \textbf{Website summary:} Please suggest an $\sim 40$ word summary for the website.Please begin with a general sentence setting the background, then outline the topics discussed in the article. You can find example summaries at \url{https://www.nature.com/natrevphys/reviews}

\section{Introduction\label{sec:intro}}
As of yet, there is no generally accepted theory of quantum gravity.
There are conceptual, mathematical, and experimental factors for this state of affairs since quantum gravity is believed to have small effects in most situations~\cite{intro1}.
Nevertheless, a variety of approaches are available which provide insight into aspects of a full theory.
Approaches can be divided into two broad categories.
The first class involves converting a classical gravity theory into a quantum version via more or less heuristic rules.
A more accurate way to put this would be to guess a quantum gravity theory based on its classical limit.
Erwin Schr\"{o}dinger came up with his famous wave equation using this method in 1926.
The rules are applicable to any gravity theory, but they are most commonly applied to general relativity.
The second class emphasizes directly constructing a fundamental quantum theory (possibly a unified theory of all interactions) from which general relativity and quantum gravity may be derived.
A primary example is the string (or M-) theory, which emerges from quantum gravity.

This paper is organized as follows. In Sect.~\ref{section2} the main theory is proposed. Then, in Sect.~\ref{section3} we examine it in non-deformed and deformed minisuperspaces. Several interesting discussions and interpretations are presented in Sect.~\ref{section4}. Finally, we conclude the results.

\section{Theory\label{section2}}
Recall the statement of Prof. Bryce S. DeWitt~\cite{dewitt}: ``In a quantum-mechanical theory of gravitation the very geometry of space and time would be subject to continual fluctuation, and even the distinction between past and future might become blurred''. The purpose of this article is to formulate this old theory of quantum gravity. It will be observed that the result of this theory leads to the fractional version of quantum gravity.

It is obvious that the nature of these fluctuations is, in general, a stochastic process.
Furthermore, it may be deduced that one of the approaches for including fluctuations into space and time is that the fluctuations may be in the heart of time $t$, and the fluctuations of the elements of configuration space feed from time.
These guide us to operational time and stochastic time arrow concepts.
In what follows, we review and formulate these concepts. Note that all the steps that will be explained below are completely principled and according to stochastic processes, not based on my own or any ad hoc assumptions. Indeed, in the following, we just translate Prof. DeWitt's statement into math.

A major characteristic of time is its direction (running from the past to the feature). This property of time will be preserved in our consideration, but we change the time clock and make it stochastic. Time in general relativity is dynamic. The first step is that we set this variable as an internal parameter $\tau$. The next step is to randomize the clock time. Every random process will not be suitable for our purpose. To save the main feature of time, it must be strictly non-decreasing. Suppose the time variable represents a sum of random intervals $T_{i}$ which can be interpreted as random waiting times. Let $T_{i}$ be independent variables that have identical distributions.
Their probability distribution does not have to be known exactly.
It is quite enough that they belong to the strict domain of attraction of an $\alpha\text{-stable}$ distribution. In order to keep the random time steps
$T_{i}$ as non-negative random variables, we need to impose a restriction: $0<\alpha<1$~\cite{FO10}.
According to~\cite{ham19}, the density $g_{\alpha}(y)$ of a positive stable random variable, like $T_{i}$, is defined by its characteristic function
\begin{align*}
\int_{0}^{\infty}\exp (j\varsigma y) g_{\alpha}(y)\mathrm{d}y
=\exp \left( -|\varsigma|^{\alpha}\exp \left[ \frac{-j\pi \alpha }{2} \mathrm{sign}(\varsigma) \right] \right),
\end{align*}
in which $j=\sqrt{-1}$.
The sum of random variables, $n^{-1/\alpha}\sum_{i=1}^{n}T_{i}, \; n \in \mathbb{N}$, converges in distribution to a $\alpha\text{-stable}$ law.
Following ref.~\cite{lya14}, under $\Delta \tau \to 0$, the following process has a limit
$T^{\Delta \tau}(\tau)=\left\{ \left\lfloor \tau / \Delta \tau \right\rfloor +1 \right\}^{-1/\alpha}\; \sum_{i=1}^{\left\lfloor \tau / \Delta \tau \right\rfloor +1 }T_{i}$, where $\tau$ is the internal time separated on discrete values with a step $\Delta \tau$, and $\left\lfloor \tau / \Delta \tau \right\rfloor$ is the integer part of $\tau / \Delta \tau$. Now we utilize the limit passage from discrete to continuous steps. As a result of the new process, the following relation is fulfilled: $T(\tau) \myeq \tau^{1/\alpha} T(1)$ where $\myeq$ means equal in distribution.
The discrete counting process is $N_{t}=\max \left\{ n \in \mathbb{N} | \sum_{i=1}^{n}T_{i} \leq t \right\}$~\cite{FO10,FO11}.
A discrete counting process' continuous limit is the first passage time or so-called hitting time which is usually denoted by $S(t)$ and is defined as $S(t)=\inf \left\{ x|T(x)>t \right\}$~\cite{lya14}. The reader who desires to know the basic properties of hitting time will do well to consult the papers~\cite{FO11, FO12}.
For a fixed time, $S(t)$ indicates the first passage of the stochastic time evolution above this time level. Note that $S(t)$ depends upon true time and it is non-decreasing hence this random process is a good option to be applied as a new time clock (stochastic time arrow).
Although $S(t)$ is self-similar, it has neither stationary nor independent increments, and all its moments are finite~\cite{lya14,lya15}.
The random process $S(t)$ is non-Markovian, but $S(T(\tau))=\tau$ meaning that $S(t)$ is inverse to the continuous limit of a Markov random process of temporal steps $T(\tau)$~\cite{FO10}. It is possible to calculate the probability density of the random variable $S(t)$ in the analytical form shown below.
Utilizing~\cite{lya14}
$$
\langle \exp(-vS(t)) \rangle=\int_{0}^{\infty}\mathcal{P}^{S}(t,x)\exp(-vx)
\mathrm{d}x=E_{\alpha}(-vt^{\alpha})
$$
where $\langle\cdots \rangle$ represents the expectation value and $E_{\alpha}$ is the Mittag-Leffler function, the probability density of the random process $S(t)$ is calculated as~\cite{FO10,ham}
\begin{align}
\mathcal{P}^{S}(t,\tau)=\frac{1}{2\pi j}\int_{\mathrm{Br}}
\exp \left( ut-\tau u^{\alpha} \right)u^{\alpha - 1}\mathrm{d}u,
\end{align}
where Br denotes the Bromwich path.
Essentially, this probability density is a way to determine the probability of being at the internal time $\tau$ on the real time $t$.
Performing the transformation $ut \to u$, the probability density would then read $\mathcal{P}^{S}(t,\tau)=t^{-\alpha}F_{\alpha}(z)$ where
\begin{align*}
F_{\alpha}=\frac{1}{2\pi j}\int_{\mathrm{Br}}\exp(u-zu^{\alpha})u^{\alpha -1}\mathrm{d}u,
\end{align*}
and $z=\tau /t^{\alpha}$.
In order to deform the Bromwich path into the Hankel path, $F_{\alpha}(z)$ can be expanded as a Taylor series~\cite{FO10,ham}:
\begin{align}
\label{tay}
F_{\alpha}(z)=\sum_{k=0}^{\infty}\frac{(-z)^{k}}{\Gamma(k+1) \Gamma(1-\alpha-k\alpha)}
\end{align}
which can be expressed in terms of H-function as
$F_{\alpha}(z)=H_{11}^{10}\left(
          \begin{array}{c}
           z \left| \begin{array}{c}
          (1-\alpha,\alpha) \\
          (0,1)
          \end{array}\right.
           \\
          \end{array}
          \right)
$~\cite{ham24}. The function $F_{\alpha}(z)$ is entire in $z$ because the radius of
convergence of the power series is infinite for $0<\alpha<1$. Hence, one can exchange between the series and the integral in the calculations of the Taylor series.
In terms of the Mittag-Leffler function, the Laplace image of the function $F_{\alpha}(z)$ can be represented~\cite{FO10}:
\begin{align*}
\int_{0}^{\infty}\exp(-z\xi)F_{\alpha}(z)\mathrm{d}z=E_{\alpha}(-\xi),
\quad z>0.
\end{align*}
For $x \geq 0$, $E_{\alpha}(-x)$ is completely monotonic if $0< \alpha \leq 1$. Furthermore, it is an entire function of order $1/\alpha$ for $\alpha >0$~\cite{lya16}. Thus, it is deduced that $F_{\alpha}(z)$ is non-negative in $z>0$.
In light of the normalization relation $\int_{0}^{\infty}F_{\alpha}(z)\mathrm{d}z=1$, it becomes readily apparent that the function $\mathcal{P}^{S}(t,\tau)$ is in fact a probability density. It is interesting to note that the Gaussian distribution $\pi^{-1/2}\exp(-z^{2}/4)$ is achieved by setting $\alpha=1/2$ in (\ref{tay}).

In quantum gravity, we deal with the Hamiltonian. Hence, now we apply the theory on Hamiltonian systems according to ref.~\cite{ham} to get basic equations.\\
Consider a Hamiltonian system whose evolution is dependent on operational time $\tau$. The corresponding equations of motion would then read
\begin{align}
\label{bcha}
\frac{\mathrm{d}q}{\mathrm{d}\tau}=\frac{\partial \mathcal{H}}{\partial p},
\qquad
\frac{\mathrm{d}p}{\mathrm{d}\tau}=-\frac{\partial \mathcal{H}}{\partial q}.
\end{align}
Clearly, for a dynamical system, each of all elements and functions, for example, $M$, which depends upon operational time, must obey such relation
\begin{equation*}
M_{\alpha}(t)=\int_{0}^{\infty}\mathcal{P}^{S}(t,\tau)M(\tau)\mathrm{d}\tau,
\end{equation*}
to convert to its fractional counterpart.
Hence, let momentum and coordinate meet the following conditions:
\begin{align*}
p_{\alpha}(t)&=\int_{0}^{\infty}\mathcal{P}^{S}(t,\tau)p(\tau)\mathrm{d}\tau,\\
q_{\alpha}(t)&=\int_{0}^{\infty}\mathcal{P}^{S}(t,\tau)q(\tau)\mathrm{d}\tau.
\end{align*}
Since $\partial \mathcal{H}/\partial q$ and $\partial \mathcal{H}/\partial p$ depend on operational time, it can be assumed that
\begin{align*}
\frac{\partial \mathcal{H}_{\alpha}}{\partial q_{\alpha}}&=\int_{0}^{\infty}\mathcal{P}^{S}(t,\tau)\frac{\partial \mathcal{H}}{\partial q}\mathrm{d}\tau,\\
\frac{\partial \mathcal{H}_{\alpha}}{\partial p_{\alpha}}&=\int_{0}^{\infty}\mathcal{P}^{S}(t,\tau)\frac{\partial \mathcal{H}}{\partial p}\mathrm{d}\tau.
\end{align*}
Therefore, in this case, equations (\ref{bcha}) become fractional:
\begin{align}
\label{bcha2}
\mathcal{D}_{t}^{\alpha}q_{\alpha}=\frac{\partial \mathcal{H}_{\alpha}}{\partial p_{\alpha}},\qquad \mathcal{D}_{t}^{\alpha}p_{\alpha}=-\frac{\partial \mathcal{H}_{\alpha}}{\partial q_{\alpha}},
\end{align}
where $\mathcal{D}_{t}^{\alpha}$ is the Caputo derivative:
\begin{align*}
\mathcal{D}_{t}^{\alpha}x(t)=\frac{1}{\Gamma (n-\alpha)}\int_{0}^{t}
\frac{x^{(n)}(\tau)}{(t-\tau)^{\alpha+1-n}}\mathrm{d}\tau,\quad n-1<\alpha <n,
\end{align*}
in which $x^{(n)}$ means the n-derivative of $x$ and $n$ is a natural number.
As a matter of fact, the fractional derivative in time is generated by the power-law waiting times.\\
For $\alpha=1$, the generalized equations (\ref{bcha2}) reduce to the ordinary Hamiltonian equations. Thus, (\ref{bcha2}) embraces a wide range of solutions.\\

\textit{Therefore, besides formulating the old theory of quantum gravity, we indicated that when we turn to fractional equations in gravity which have recently been addressed (for example see refs.~\cite{Fr1,Fr2,Fr3,Fr4,Fr5,Fr6,Fr7}), we are actually considering stochastic perturbations of space-time implicitly, in other words, it was demonstrated that fractional gravity is generated by stochastic fluctuations of space-time.}

\section{An example for application\label{section3}}
In this section, we want to apply the formulation made in the previous section to quantum gravity.

\subsection{The model in non-deformed minisuperspace\label{sec:intro}}
Observations indicate a flat homogeneous and isotropic universe at a large scale. Hence, we start with the FLRW metric of the form
\begin{align}
\mathrm{d}s^2=-N(t)^2 \mathrm{d}t^2+a(t)^2 \left [\mathrm{d}x^2+\mathrm{d}y^2
+\mathrm{d}z^2 \right],
\end{align}
where $N(t)$ and $a(t)$ are the lapse function and scale factor of the universe, respectively. One of the fundamental theories of gravity is the Einstein-Hilbert action along with a homogeneous scalar field $\varphi(t)$ dominating over matter degrees of freedom. Therefore, our action may take the form
\begin{align}
S=\int \mathrm{d}t \left\{ \frac{-3a \dot{a}^2}{N}+a^3 \left( \frac{ \dot{\varphi}^2}{2N} -N \Lambda\right) \right\},
\end{align}
by restricting ourselves to constant potential $\Lambda$ (cosmological constant). Here, the dot denotes a differentiation with respect to time $t$.
Now we apply the aforementioned theory. Hence $a$ and $\varphi$ are mapped to fractional version $a_{\alpha}$ and $\varphi_{\alpha}$, respectively.
Since we deal with Hamiltonian, the simple form of it is advantageous. In order to procure this, let us perform the following change of variables
\begin{align}
\left\{
\begin{array}{ll}
q_{\alpha x}&=\lambda^{-1} a_{\alpha}^{3/2} \sinh \left( \lambda \varphi_{\alpha} \right);\nonumber \\[2\jot]
q_{\alpha y}&=\lambda^{-1} a_{\alpha}^{3/2} \cosh \left(\lambda \varphi_{\alpha} \right),
\end{array}
\right.
\end{align}
where $\lambda^{-1}=\sqrt{8/3}$.
Utilizing the above transformations, the (fractional) canonical Hamiltonian (first class constraint) would then read
\begin{align}
\mathcal{H}_{\alpha c.}=N \left(\frac{1}{2}p_{\alpha x}^{2}+\frac{\omega^{2}}{2} q_{\alpha x}^2 \right)
- N \left(\frac{1}{2}p_{\alpha y}^{2}+\frac{\omega^{2}}{2} q_{\alpha y}^2 \right),
\end{align}
in which $\omega^{2}=-3 \Lambda /4$. As is observed, the Hamiltonian appeared as a fractional ghost oscillator namely as a difference of two fractional harmonic oscillators.\\
The elements of our new configuration space, $(q_{\alpha x}\, , \, q_{\alpha y})$, and their conjugate momentums fulfill the following relations\footnote{
Note that $k$ and $j$ can take $1$ and $2$, i.e. $(x_{1},x_{2})=(x,y)$.}:
\begin{align}
&\{ q_{\alpha x_{k}},q_{\alpha x_{j}}\}=0, \quad \{ p_{\alpha x_{k}}, p_{\alpha x_{j}} \}=0, \nonumber\\&
\{ q_{\alpha x_{k}}, p_{\alpha x_{j}} \}= \delta_{kj},
\end{align}
where $\{ \cdots \}$ is the Poisson bracket and $\delta_{kj}$ is the usual Kronecker delta.\\
The equations of motion indicating the dynamics of our system are:
\begin{align}
&\mathcal{D}_{t}^{\alpha} q_{\alpha x}=p_{\alpha x},
\quad
\mathcal{D}_{t}^{\alpha} p_{\alpha x}=-\omega^{2}q_{\alpha x},\\
&\mathcal{D}_{t}^{\alpha} q_{\alpha y}=-p_{\alpha y},
\quad
\mathcal{D}_{t}^{\alpha} p_{\alpha y}=\omega^{2}q_{\alpha y},
\end{align}
yielding
\begin{align}
\mathcal{D}_{t}^{2 \alpha} q_{\alpha x}+\omega^{2} q_{\alpha x}=0,\\
\mathcal{D}_{t}^{2 \alpha} q_{\alpha y}+\omega^{2} q_{\alpha y}=0.
\end{align}
Solving this fractional system using the Laplace approach gives
\begin{align}
\label{001}
q_{\alpha x}(t) &=c_{1} \cosh_{\alpha}\left( \left| \omega \right| t^{\alpha} \right)
+c_{2} \sinh_{\alpha}\left( \left| \omega \right| t^{\alpha} \right),\\
\label{002}
q_{\alpha y}(t) &=c_{3} \cosh_{\alpha}\left( \left| \omega \right| t^{\alpha} \right)
+c_{4} \sinh_{\alpha}\left( \left| \omega \right| t^{\alpha} \right),
\end{align}
where $c_{j}$s are constants of integration and $\sinh_{\alpha}$ and $\cosh_{\alpha}$ are fractional-hyperbolic sine and cosine functions:
\begin{align}
\sinh_{\alpha}(z)&=\frac{E_{\alpha}(z)-E_{\alpha}(-z)}{2},\\
\cosh_{\alpha}(z)&=\frac{E_{\alpha}(z)+E_{\alpha}(-z)}{2},
\end{align}
in which $E_{\alpha}$ is the Mittag--Leffler function.\\
\begin{figure}[ht]
\centering
\includegraphics[width=\columnwidth]{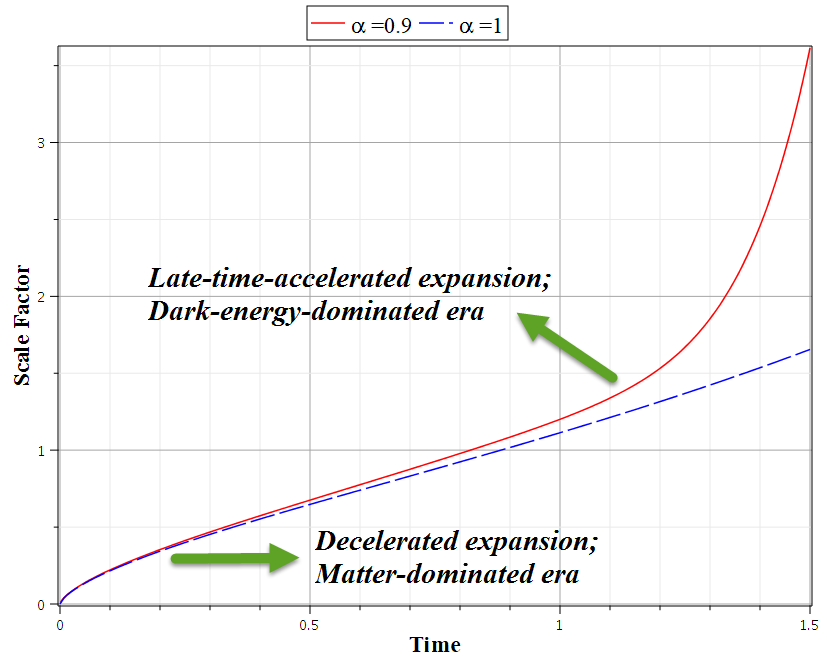}\\
\caption{The solid curve indicates the behavior of scale factor (\ref{scfa}) for the values $\alpha=0.9$, $(c_{4}^{2}- c_{2}^{2})=1$, $\lambda=1$, and $\omega=1$ while the dashed curve demonstrates the plot for $\alpha=1$ (limiting case; usual quantum gravity).}\label{fig1}
\end{figure}
Sinh carries both decelerated and accelerated eras while cosh corresponds to only an accelerated era. In the case of preserving both cosh and sinh, the transition from a decelerated expansion to an accelerated epoch will only be possible when cosh's amplitude is incredibly small compared to sinh's.
As an example, by assuming $c_{1}\simeq c_{3}\simeq 0$, the scale factor of the universe becomes
\begin{align}\label{scfa}
a_{\alpha}(t)\simeq(c_{4}^{2}- c_{2}^{2})^{1/3} \left[ \lambda \sinh_{\alpha}(|\omega | t^{\alpha})\right]^{2/3}.
\end{align}
The qualitative behavior of this scale factor has been plotted in fig.~(\ref{fig1}).
As is observed, it demonstrates a transition from decelerated to accelerated expansion
(late-time-accelerated expansion).
Many astrophysical observations have verified this behavior of our universe, including supernova type Ia~\cite{ph1,ph2}, large scale structure~\cite{ph5}, weak lensing~\cite{ph4}, CMB studies~\cite{ph3}, and baryon acoustic oscillations~\cite{ph6}. Moreover, the expansion obtained from the fractional equations demonstrated more flexibility than those of usual quantum gravity and represented the ``super-accelerated'' expansion for the current status of the universe.

By comparing (\ref{001})-(\ref{002}) and also (\ref{scfa}) with the results of \cite{behplb}, it becomes clear that (\ref{001})-(\ref{002}) and (\ref{scfa}) are more general because for $\alpha=1$ they reproduce the results of \cite{behplb} (Correspondence principle).\\

\S \;\; \textbf{An alternative way of getting solutions: Fractional normal modes}\\
Like the ladder operators (raising and lowering operators) of the harmonic oscillators in quantum theory, here we may also define the following objects:
\begin{align}
A_{x_{j}}(t)&= \sqrt{\frac{\omega}{2}} \left( q_{x_{j}} +\frac{i \, p_{x_{j}}}{\omega} \right),\\
A_{x_{j}}^{\star}(t)&= \sqrt{\frac{\omega}{2}} \left( q_{x_{j}} -\frac{i \, p_{\alpha x_{j}}}{\omega} \right).
\end{align}
Now by randomizing the time clock of the values $A_{x_{j}}(\tau)$ and $A_{x_{j}}^{\star}(\tau)$ so that the characteristic time is absent and then determining new objects as
\begin{align}
A_{\alpha x_{j}}(t)&= \int_{0}^{\infty} \mathcal{P}^{S}(t ,\tau) A_{x_{j}}(\tau) \mathrm{d}\tau, \label{ladderequs1}\\
A_{\alpha x_{j}}^{\star}(t)&= \int_{0}^{\infty} \mathcal{P}^{S}(t ,\tau) A_{x_{j}}^{\star}(\tau) \mathrm{d}\tau, \label{ladderequs2}
\end{align}
one gets
\begin{align}
A_{\alpha x_{j}}(t)&= \sqrt{\frac{\omega}{2}} \left( q_{\alpha x_{j}} +\frac{i \, p_{\alpha x_{j}}}{\omega} \right),\\
A_{\alpha x_{j}}^{\star}(t)&= \sqrt{\frac{\omega}{2}} \left( q_{\alpha x_{j}} -\frac{i \, p_{\alpha x_{j}}}{\omega} \right).
\end{align}
Here, $A_{\alpha x_{j}}^{\star}(t)$ is the adjoint of $A_{\alpha x_{j}}(t)$.\\
These relations can be inverted, leading to
\begin{align}
q_{\alpha x_{j}}&=\frac{1}{\sqrt{2\omega}} \left(A_{\alpha x_{j}}(t)+A_{\alpha x_{j}}^{\star}(t)\right), \label{ladderequs3} \\
p_{\alpha x_{j}}&=-i \, \sqrt{\frac{\omega}{2}} \left( A_{\alpha x_{j}}(t) - A_{\alpha x_{j}}^{\star}(t) \right).
\end{align}
Inserting the solutions of usual normal modes of harmonic oscillators which obey the equations
\begin{align}
\dot{A}_{x_{j}}(t)&=-i \omega A_{x_{j}}(t),\\
\dot{A}_{x_{j}}^{\star}(t)&=i \omega A_{x_{j}}^{\star}(t),
\end{align}
in~(\ref{ladderequs1})--(\ref{ladderequs2}) and then putting in~(\ref{ladderequs3}) and performing some computations lead exactly to what we obtained utilizing the Laplace approach ((\ref{001})-(\ref{002})).

\subsection{The model in deformed minisuperspace\label{sec:intro}}
The concept of deformed minisuperspace, also called deformed phase space, was first introduced with non-commutative cosmology~\cite {prd03}.
They introduced a deformation of the minisuperspace to accommodate the non-commutativity in order to avoid the complications related to the non-commutative theory of gravity.

After canonical quantization, the Wheeler-DeWitt equation is formally obtained in canonical quantum cosmology. This is a Klein-Gordon-type equation that the quantum behavior of the Universe can be described by it. Deformations in the phase space of a system can be used as an alternative approach to studying quantum mechanical effects.
The approach is an equivalent path to quantization and is part of a complete and consistent type of quantization known as deformation quantization~\cite{prd07}.
It can be assumed that investigating cosmological models in deformed minisuperspace could be interpreted as studying quantum effects to cosmological solutions~\cite{prd08}.
Formally, in the deformed minisuperspace approach, we introduce the deformation by Moyal brackets
\begin{align*}
\{F,G\}_{\nu}=F\star_{\nu}G - G\star_{\nu}F,
\end{align*}
in which the product between $F$ and $G$ is replaced by Moyal product
\begin{align}
\label{star0}
\left( F\star G \right)(x)=\exp\left[\frac{1}{2}\nu^{ab}\partial_{a}^{(1)}
\partial_{b}^{(2)} \right]\left.F(x_{1})G(x_{2}) \right|_{x_{1}=x_{2}=x}
\end{align}
such that
\begin{align*}
\nu = \left(
        \begin{array}{cc}
          \theta_{ij} & \delta_{ij}+\sigma_{ij} \\
          -\delta_{ij}-\sigma_{ij} & \beta_{ij} \\
        \end{array}
      \right)
\end{align*}
where $\theta_{ij}$ and $\beta_{ij}$ are $2\times 2$ antisymmetric matrices and they represent the non-commutativity in the coordinates and momenta, respectively. But in our case, the partial derivatives (bidifferential operators) are substituted with fractional derivatives. Below is the algebra resulting from $\nu$ deformed algebra for the variables in minisuperspace:
\begin{align}
\label{num01}
&\left\{q_{\alpha x_{i}},q_{\alpha x_{j}}\right\}_{\nu}=\theta_{ij},
\qquad
\left\{p_{\alpha x_{i}},p_{\alpha x_{j}}\right\}_{\nu}=\beta_{ij}, \nonumber\\&
\left\{q_{\alpha x_{i}},p_{\alpha x_{j}}\right\}_{\nu}=\delta_{ij}+\sigma_{ij}
\end{align}
Like ref.~\cite{behplb}, here we investigate particular expressions for the deformations as $\theta_{ij}=-\theta \epsilon_{ij}$ and $\beta_{ij}=\beta \epsilon_{ij}$ where $\epsilon_{ij}$ is the two dimensional Levi-Civita tensor.

We can derive a similar algebra to (\ref{num01}) utilizing an alternative method as follows.
There will be no difference in the algebra, but the Poisson brackets will be different in the two algebras.
For Eq. (\ref{num01}), the brackets are $\nu$ deformed ones and are related to the Moyal product, and for the other algebra the brackets are the usual Poisson brackets.
Taking the classical phase space variables and performing the following transformation
\begin{align}
\left(\begin{array}{c}
  \hat{q}_{\alpha x} \\
  \hat{q}_{\alpha y}
\end{array}
\right)
&=
\left(\begin{array}{c}
  q_{\alpha x} \\
  q_{\alpha y}
\end{array}\right)
+\frac{\theta}{2}
\left(\begin{array}{c}
  p_{\alpha y} \\
  -p_{\alpha x}
\end{array}\right),
\\
\left(\begin{array}{c}
  \hat{p}_{\alpha x} \\
  \hat{p}_{\alpha y}
\end{array}
\right)
&=
\frac{\beta}{2}\left(\begin{array}{c}
  -q_{\alpha y} \\
  q_{\alpha x}
\end{array}\right)
+
\left(\begin{array}{c}
  p_{\alpha x} \\
  p_{\alpha y}
\end{array}\right),
\end{align}
the algebra now reads as follows:
\begin{align}
&\left\{\hat{q}_{\alpha y},\hat{q}_{\alpha x}\right\}=\theta,\qquad
\left\{\hat{p}_{\alpha y},\hat{p}_{\alpha x}\right\}=\beta,\\
&\left\{\hat{q}_{\alpha x},\hat{p}_{\alpha x}\right\}=1+\sigma,\qquad
\left\{\hat{q}_{\alpha y},\hat{p}_{\alpha y}\right\}=1+\sigma,
\end{align}
in which $\sigma=\theta \beta /4$.
For the rest of this paper, we will use this modified algebra.\\
Now, the Hamiltonian based on the variables that obey the modified algebra takes the form:
\begin{align}
\mathcal{H}_{\alpha\mathrm{,def.}}=N \left(\frac{1}{2}\hat{p}_{\alpha x}^{2}+\frac{\omega^{2}}{2} \hat{q}_{\alpha x}^2 \right)
-N \left(\frac{1}{2}\hat{p}_{\alpha y}^{2}+\frac{\omega^{2}}{2} \hat{q}_{\alpha y}^2 \right),
\end{align}
or equivalently
\begin{align}
&\mathcal{H}_{\alpha\mathrm{,def.}}=\nonumber\\&\frac{1}{2} \bigg(
p_{\alpha x}^{2}-p_{\alpha y}^{2} -w_{1}^{2}
\left( x p_{\alpha y}+y p_{\alpha x} \right)
+w_{2}^{2} \left( x^2 -y^2 \right)
\bigg),
\end{align}
where
\begin{align}
w_{1}^{2}=\frac{4 \beta-4 \omega^{2} \theta}{4- \omega^{2} \theta^{2}}, \qquad w_{2}^{2}= \frac{4\omega^{2}- \beta^{2}}{4- \omega^{2} \theta^{2}}.
\end{align}
The equations of motion are then obtained as
\begin{align}
&\mathcal{D}_{t}^{\alpha}q_{\alpha x}=p_{\alpha x}
-\frac{1}{2}w_{1}^{2}q_{\alpha y},\quad
\mathcal{D}_{t}^{\alpha}p_{\alpha x}=\frac{1}{2}w_{1}^{2}p_{\alpha y}
-w_{2}^{2}q_{\alpha x},\\
&\mathcal{D}_{t}^{\alpha}q_{\alpha y}=-p_{\alpha y}
-\frac{1}{2}w_{1}^{2}q_{\alpha x},\quad
\mathcal{D}_{t}^{\alpha}p_{\alpha y}=\frac{1}{2}w_{1}^{2}p_{\alpha x}
+w_{2}^{2}q_{\alpha y},
\end{align}
yielding
\begin{align}
\mathcal{D}_{t}^{2\alpha}q_{\alpha x}+w_{1}^2 \mathcal{D}_{t}^{\alpha}q_{\alpha y}
+w_{3}^{2} q_{\alpha x}=&0, \label{defeqs01} \\
\mathcal{D}_{t}^{2\alpha}q_{\alpha y}+w_{1}^2 \mathcal{D}_{t}^{\alpha}q_{\alpha x}
+w_{3}^{2} q_{\alpha y}=&0,\label{defeqs02}
\end{align}
in which $w_{3}^{2}=\left( 4w_{2}^{2}+\epsilon w_{1}^{2} \right)/4$.
In order to solve this system, let us first decouple this coupled system through the transformations:
\begin{align*}
\left\{
  \begin{array}{ll}
   q_{\alpha x}=X_{\alpha 1}+X_{\alpha 2}; \\[2\jot]
   q_{\alpha y}=X_{\alpha 1} - X_{\alpha 2}.
  \end{array}
\right.
\end{align*}
Therefore, (\ref{defeqs01})--(\ref{defeqs02}) become
\begin{align}
\mathcal{D}_{t}^{2\alpha}X_{\alpha 1}+2 \gamma_{1} \mathcal{D}_{t}^{\alpha}X_{\alpha 1}
+w_{3}^{2} X_{\alpha 1}=&0, \label{defeqs03} \\
\mathcal{D}_{t}^{2\alpha}X_{\alpha 2}+2 \gamma_{2} \mathcal{D}_{t}^{\alpha}X_{\alpha 2}
+w_{3}^{2} X_{\alpha 2}=&0, \label{defeqs04}
\end{align}
in which $\gamma_{1}=w_{1}^{2}/2$ and $\gamma_{2}=-\gamma_{1}$.
Both of these equations may be named fractional damped harmonic oscillators because for $\alpha=1$ they reduce to the usual equation of damped harmonic oscillators of classical mechanics.\\
Like classical mechanics, there are three possibilities for the fractional damped oscillators, but since we want to compare our results with ref.~\cite{behplb}, here we also focus on late-time-accelerated expansion and the phase transition and present just one solution.\\
First of all, let us define the function:
\begin{align}
  &\Psi\left(p,q \left| t^{\alpha} \right. \right)=\nonumber\\&
\sum_{n,m=0}^{\infty}\frac{\Gamma(2n+m+2)\,p^{m}\,q^{2n+1}\,t^{(2n+m+1)\alpha}}
  {\Gamma((2n+m+1) \alpha+1 )\,\Gamma(2n+2)\,\Gamma(m+1)},
\end{align}
which at the limit $\alpha=1$ leads to $\exp(pt)\sinh(qt)$.\\
One of the solutions of equations of motion is as
\begin{align}
q_{\alpha x}(t)&=C_{1}\Psi_{2}
+C_{2}\Psi_{1},\\
q_{\alpha y}(t)&=C_{1}\Psi_{2}
-C_{2}\Psi_{1},
\end{align}
in which $C_{1}$ and $C_{2}$ are constants of integration and $\Psi_{1}=\Psi\left(\gamma_{1}, w_{4} \left| t^{\alpha} \right. \right)$, and $\Psi_{2}=\Psi\left(\gamma_{2}, w_{4} \left| t^{\alpha} \right. \right)$ where $w_{4}=\sqrt{-w_{2}^{2}}$.
Consequently, the scale factor of the Universe in the `C' and `NC' frames are respectively given by
\begin{align}
a_{\alpha, \mathrm{C}}(t)=& \left\{-4\lambda C_{1}C_{2}\Psi_{1}
\Psi_{2}\right\}^{1/3},\\
a_{\alpha, \mathrm{NC}}(t)=&\bigg\{\frac{-\lambda C_{1}C_{2}}{4}
\left[ 2\theta
\Phi_{2}
+\left( \theta w_{1}^{2}-4 \right)\Psi_{2}
\right]\nonumber\\&\times
\left[ -2\theta
\Phi_{1}
+\left( \theta w_{1}^{2}-4 \right)\Psi_{1}
\right]\bigg\}^{1/3},
\end{align}
where
$\Phi_{1}=\Phi\left(\gamma_{1}, w_{4} \left| t^{\alpha} \right. \right)$, and $\Phi_{2}=\Phi\left(\gamma_{2}, w_{4} \left| t^{\alpha} \right. \right)$ in which
\begin{align}
  \Phi\left(p,q \left| t^{\alpha} \right. \right)
  =\sum_{n,m=0}^{\infty}\frac{\Gamma(2n+m+2)\,p^{m}\,q^{2n+1}\,t^{(2n+m)\alpha}}
  {\Gamma((2n+m) \alpha+1 )\,\Gamma(2n+2)\,\Gamma(m+1)}.
\end{align}
The qualitative behavior of these scale factors for small values of $\theta$ are similar to fig.\ref{fig1}; for example, for the selections $\alpha=0.9$,
$\theta=0.001$, $\gamma_{1}=1$,
$\gamma_{2}=-1$, $w_{4}=1$, $w_{1}=1$, $\lambda C_{1}C_{2}=-4$ we
get the same behavior.\\
When $\alpha$ approaches $1$ the above solutions become
\begin{align}
a_{\alpha,\mathrm{C}}(t)&=\left\{ -4\lambda C_{1}C_{2} \sinh^{2}(w_{4}t) \right\}^{1/3},\label{cosf1}
\end{align}
\begin{align}
\label{after2}
a_{\alpha,\mathrm{NC}}(t)=\left\{ \lambda C_{1} C_{2} \left[ -4\sinh^{2}(w_{4}t)+\theta^{2}w_{4}^{2}\cosh^{2}(w_{4}t) \right] \right\}^{1/3},
\end{align}
which coincide with the results of ref.~\cite{behplb}.\\
Therefore, we conclude that by the use of the stochastic time arrow theory in both non-deformed and deformed phase spaces, we obtain generalized solutions that in the limit $\alpha=1$, reduce to those of usual quantum gravity. In fact, chaos manifests itself through $\alpha$.

\section{Indirectly proving the fluctuations and interesting discussions and interpretations\label{section4}}
In this section, some interesting discussions and interpretations are presented.\\

$\bullet$ \textbf{Conservation of energy:} According to ref.~\cite{F.O.}, the fractional oscillator can be considered as an ensemble average of harmonic oscillators. The full contribution of all harmonic oscillators is equal to the product of their number and the response of one oscillator when all oscillators are identical and running in the same phase. This occasion corresponds to $\alpha=1$.
Oscillators that differ a little in frequency from each other, even if they are phase-synchronized at the start point, are eventually distributed uniformly across the clock face.
In such a system ($\alpha \neq 1$), each response will be counterphased by another oscillator, resulting in a compensated response from all harmonic oscillators.
In spite of the fact that each oscillator is conservative (its total energy is saved), the system of these oscillators (i.e. fractional oscillator) results in a dissipative nature.
In our case, for example in a non-deformed space, a fractional oscillator is attributed to each of `$x$' and `$y$', each of which is damped according to the discussion that took place, but we can take their amplitudes as `$+A$' and `$-A$' to neutralize each other's effect and maintain the total zero energy. In gravity, the minimum number of fractional oscillators is two; one comes from gravity (in our case $a_{\alpha}$ ) and one esteemed from matter content (in our case $\varphi_{\alpha}$). Hence, tuning the amplitudes to get zero total energy is straightforward. Therefore, despite existing fractional oscillators, the conservation of energy is attainable.\\

$\bullet$ \textbf{Why fractional system?} Our system has two subsystems: $x$ and $y$ which esteemed from the elements of configuration space $a_{\alpha}$ and $\varphi_{\alpha}$. The whole system is isolated hence $a_{\alpha}$ can be regarded as an environment for $\varphi_{\alpha}$ and vice versa. It is conditioned upon the interaction between subsystems and the environment that results in a modification of the conventional representation of the Hamilton equations. It is each subsystem's internal clock that governs its behavior. However, its coordinates and momenta depend on the operational time, despite the Hamiltonian equations describing its dynamics. Averaging accounts for the interaction between the subsystem and environment during the passage from operational to physical time~\cite{ham}. Consequently, both subsystems behave as a fractional system.\\

$\bullet$ \textbf{A step towards making the system real:}
The solutions (\ref{001})-(\ref{002}) can be recast in the form
$C_{1}F_{1}(t)+C_{2}F_{2}(t)$
where $F_{1}(t)=E_{2\alpha , 1}\left( -\omega^{2} t^{2\alpha} \right)$ and $F_{2}(t)=\omega t^{\alpha}
E_{2\alpha , \alpha +1}\left( -\omega^{2} t^{2\alpha} \right)$ in which $E_{\mu , \nu}$ is the two parameter Mittag-Leffler function. The value $F_{1}(t)$ satisfies the equation~\cite{F.O.}
\begin{align}
\label{pk}
F_{1}(t)=1-\frac{\omega^{2}}{\Gamma (2\alpha)}\int_{0}^{t}
\left( t- t^{\prime}\right)^{2\alpha -1}F_{1}\left(t^{\prime} \right)\mathrm{d}t^{\prime}.
\end{align}
It is also possible to write the appropriate equation for $F_{2}(t)$. According to ref.~\cite{F.O.}, the power kernel in (\ref{pk}) interpolates the memory function between the Dirac function (the absence of memory) and the step function (complete ideal memory) meaning that memory manifests itself within all the time interval. Note that the kernels of usual solutions of quantum gravity like those that exist in \cite{behplb} behave like a complete ideal memory ($\alpha =1$) in which the system remembers all its states, and this excites the harmonic oscillations in such a system. Neither the absence of memory that causes relaxation is correct nor complete ideal memory. It seems that a real system inherits the sum of both natures which is what fractional solutions do. This is similar to having a rope tied at both ends and we create a wave in it. This rope is not indifferent to the effect of this wave, nor does it continue to oscillate forever. Depending on the conditions, the rope oscillates for a while and then returns to its original state.\\
Hence, our solutions may be regarded as a step towards making the system real.
\\
\textit{This discussion and the previous one may be regarded as indirect proof for existing fluctuations in space-time structure. It seems that the experimental proof, like the proof of the string theory, requires high energies.}

$\bullet$ \textbf{Causality principle and closed time-like curves:} Since the direction of time was preserved in our consideration, hence the causality principle is automatically maintained.
From the existence of the principle of causality on the one hand and the stochastic nature of the studied process on the other hand, it is felt that the presence of closed time-like curves and time travel is forbidden from our vantage point and hence the process (evolution of the universe) is irreversible.

\section{Conclusion}
``In quantum gravity, the very geometry of space and time would be subject to continual fluctuation.'' Prof. Bryce Seligman DeWitt said. In this paper, we translated this statement into mathematical language.
In this formulation, we considered fluctuations of a general type, i.e. stochastic, by maintaining the accepted assumptions for the time such as its direction. This upgrade implied that equations be shifted to fractional order. Indeed, when one wants to include fluctuations, it is sufficient that the Hamiltonian equations of usual quantum gravity be converted to corresponding fractional order. It means that if the order of a differentiation operator is $n$, then the order of the corresponding fractional element shall be $\nu$ where $n-1<\nu \leq n$.
Fractional gravity have recently received a lot of attention, leading to graceful results. The authors apply this type of gravity because it works, not for any deep reason. This paper provided a solid foundation for using fractional gravity: ``Fractional quantum gravity arises from stochastic fluctuations in space-time''.\\
The existence of such fluctuations in the space-time was proved by an indirect approach.\\
%Our paper indicated that in these articles, by shifting to fractional equations, the authors have actually considered space-time fluctuations implicitly.\\
For clarification, we presented an example of applying the theory in a well-known Einstein-Hilbert theory along with a scalar field in deformed and non-deformed minisuperspaces. The final results indicated generalized solutions which in a certain limit reproduce the results of usual quantum gravity and hence the correspondence principle is guaranteed.
%The mentioned certain limit was $\alpha=1$ where $\alpha$ is the order of fractional equations.
All of our solutions indicated a transition from decelerated to accelerated expansion (late-time-accelerated expansion) which conforms with observations. Moreover, the expansion obtained from the fractional equations demonstrated more flexibility than those of usual quantum gravity and represented the ``super-accelerated'' expansion for the current state of the universe.
Finally, exciting discussions were performed regarding energy conservation, the reason for leading to fractional systems, the causality principle, and the existing ban on closed time-like curves. Moreover, a justification that fractional equations are a step towards making the system real was presented.

%\acknowledgments
%

%\section*{Author contributions}
%All aspects of the work have been carried out by B. Tajahmad.

%\section*{Competing interests}
%The authors declare no competing interests. 

%\section*{Publisher’s note}
%Springer Nature remains neutral with regard to jurisdictional claims in published maps and institutional affiliations.

%\section*{Supplementary information (optional)}
%If your article requires supplementary information, please include these files for peer-review. Please note that supplementary information will not be edited.

%\newpage
%\section*{Box 1 (Optional)}
%This is a Box, which can contain a figure, and which should have no more than 300 words of text.

%\begin{figure}[ht]
%\centering
%\includegraphics[width=\linewidth]{fig}
%\caption{The figure caption should start with a title explaining the figure. Figures should be self-consistent so please redefine all acronyms and define all symbols. Example: GHZ, Greenberger–Horne–Zeilinger, OAM, orbital angular momentum. Please provide credit lines for panels reproduced from the literature. Example: Panels a and b are reproduced from Ref. \cite{TR}.}
%\label{fig}
%\end{figure}

\end{document}